# *Post-transcriptional knowledge in pathway analysis increases the accuracy of phenotypes classification*

# Post-transcriptional knowledge in pathway analysis increases the accuracy of phenotypes classification

Salvatore Alaimo[1], Rosalba Giugno[2], Mario Acunzo[3], Dario Veneziano[3], Alfredo Ferro[2], and Alfredo Pulvirenti[2,*]

[1]Department of Mathematics and Computer Science and [2]Department of Clinical and Experimental Medicine, University of Catania, Catania, Italy,
[3]Department of Molecular Virology, Immunology and Medical Genetics, Comprehensive Cancer Center, The Ohio State University, Columbus, OH, USA.

## Abstract

**Motivation:** Prediction of phenotypes from high-dimensional data is a crucial task in precision biology and medicine. Many technologies employ genomic biomarkers to characterize phenotypes. However, such elements are not sufficient to explain the underlying biology. To improve this, pathway analysis techniques have been proposed. Nevertheless, such methods have shown lack of accuracy in phenotypes classification.

**Results:** Here we propose a novel methodology called *MITHrIL* (Mirna enrIched paTHway Impact anaLysis) for the analysis of signaling pathways, which has built on top of the work of Tarca et al., 2009. *MITHrIL* extends pathways by adding missing regulatory elements, such as microRNAs, and their interactions with genes. The method takes as input the expression values of genes and/or microRNAs and returns a list of pathways sorted according to their deregulation degree, together with the corresponding statistical significance (p-values). Our analysis shows that *MITHrIL* outperforms its competitors even in the worst case. In addition, our method is able to correctly classify sets of tumor samples drawn from TCGA.

**Availability:** MITHrIL is freely available at the following URL: http://alpha.dmi.unict.it/mithril/

## Introduction

The prediction of phenotypes, such as diseases, or of responses to therapies from the large amount of genotypic high-dimensional data obtained through *Next-Generation Sequencing* techniques is an extremely important task in translational biology and precision medicine. However, the gap between current analysis techniques and the ability to obtain accurate knowledge is broad.

High-throughput sequencing and gene profiling techniques are radically transforming medical research, allowing the full monitoring of a biological system. The use of these technologies typically generates a list of differentially expressed elements (i.e. genes or microRNAs) whose behavior varies significantly among the phenotypes under examination.

Furthermore, compared to traditional gene expression extraction techniques (eg. Microarray), deep sequencing methods, such as RNA-Seq, provide much larger lists of differentially expressed genes, increasing, therefore, the complexity of the analysis. The common approach to simplify and make the analysis of such data more fruitful consisted in grouping genes into smaller sub-sets according to some relationship, leveraging on existing knowledge-bases such as ontologies or pathways. The analysis of this type of data at the functional level is crucial since it allows a strong reduction of dimensionality, thus providing greater insights on the biology of the phenomenon under study [1].

---

[*] To whom correspondence should be addressed.

An extensive class of techniques known as *Pathway Analysis* [2] goes in this direction. In the past, such term has been associated to the analysis of ontological terms, protein-protein interaction (PPI) networks, or to the inference of gene regulatory networks from expression data. More recently, great interest has shifted toward a class of methods called *Knowledge base-driven pathway analysis* [3]. Such methods leverage on existing databases, such as the Kyoto Encyclopedia of Gene and Genomes (KEGG) [4, 5] or Pathway Commons [6], to identify those pathways that may be affected by the expression changes in the observed phenotype. Knowledge base-driven pathway analysis techniques can be grouped into three generations of approaches [3]: i) Over-Representation Analysis (ORA); ii) Functional Class Scoring (FCS); iii) Pathway Topology-based (PT).

First-generation methods statistically evaluate the number of altered genes in a pathway with respect to the set of all analyzed genes. After filtering the resulting gene set of an expression assessment experiment, ORA strategies [7-13] typically divide the list of genes according to the pathway each gene belongs to. By applying an hypothesis test (i.e. hypergeometric, chi-square, or binomial) they are able to determine if the number of such genes is over- or under-represented. These methods, however, have some major limitations. Firstly, considering only the number of differentially expressed genes, while omitting their expression, implies that the magnitude of their change be unimportant for pathway activity. Furthermore, considering only statistically significant differential expression may exclude those genes whose coordinated alteration may lead to remarkable effects, although their differential expression may not be statistically significant. Finally, they consider individual genes and pathways, respectively, in an manner independent of the surrounding biological context, eluding what truly happens in reality.

A first generation method, DIANA-miRPath [14], assesses the impact of miRNAs in biological processes by identifying the pathways in which they are significantly involved. The software employs a functional annotation of one or more miRNAs by means of a hypergeometric distribution, or an unbiased empirical distribution, or a statistical meta-analysis. Moreover, it allows the identification of sub-sets of miRNAs which significantly regulates a collection of pathway, on the basis of experimental data.

Second-generation methods compensated some of the disadvantages of ORA approaches. Typically FCS methods compute a gene-level statistic from their expression levels, by means of a statistical approach (i.e. ANOVA, Q-statistic, signal-to-noise ratio, t-test, or Z-score). Such a statistic is calculated considering all genes in a pathway [15-21] and its statistical significance is estimated through an appropriate null hypothesis [16, 22-24]. FCS methods avoid some of the limitations of the ORA approaches by ranking all genes through their expression level and by considering the dependencies within a pathway. However, by using gene expression only to sort genes, they do not take into account the fact that such values can change pathways activity.

In order to overcome the disadvantages of FCS methods, the third class of techniques models a pathway as a graph, considering its topology when computing scores. A thorough analysis of all PT-based approaches has been provided in [25].

In Draghici et al. [26], an analytical technique called *impact factor* (*IF*) was introduced. The impact factor is a pathway-level score that takes into account biological factors such as the magnitude of change in genes expression, the type of interactions between genes, and their location in the pathway. In Draghici et al. [26], each pathway is modeled as a graph in which nodes represent genes, while edges represent interactions between them. Authors also define a gene-level statistic (called *perturbation factor*, *PF*) as a linear function of the change in gene expression and the perturbation of its neighborhood. Such a statistic is then combined for each element in a pathway, and a p-value is computed by means of exponential distribution.

The analysis method presented by Draghici et al. [26], has been further improved by the *SPIA* algorithm [27] which attenuates the dominant effect exercised by the change in expression within *PFs* computation, while reducing the high rate of false positives when the input list of genes is small. *SPIA* uses a bootstrap procedure to evaluate the significance of the observed perturbation in the pathway. All this is combined with a p-value computed in ORA style to make a full assessment of the statistical significance of the perturbation of each pathway.

To reduce the number of false positives, and to obtain a more significant analysis, Vaske et al. [28] presented the *PARADIGM* algorithm, which has been further improved by [29]. *PARADIGM* is a method to infer patient-specific genetic activity by incorporating information regarding interactions between genes provided in a pathway. The method predicts the degree of alteration in the activity of a pathway by employing a probabilistic inference algorithm. The authors show that their model obtains significantly more reliable results than *SPIA*. However, Mitrea et al. [25] stated they could not reproduce the results reported in Vaske et al. [28], despite the full cooperation of its authors.

However, both SPIA and PARADIGM completely ignore post-transcriptional regulatory interactions enacted by miRNAs. To fill this gap, Calura et al. developed a new approach, Micrographite [30], which is able to integrate pathway with predicted and validated miRNA-target interactions. The method, by performing a topological analysis based on expression profiles, is able to identify significant gene circuits specific of a phenotype. The main advantage of the methodology is the ability to accurately describe the cellular scenario that led to the input expression data.

Here, we present *MITHrIL* (miRNA enriched pathway impact analysis), a technique that extends the method in [26] and *SPIA* [27], by combining their effectiveness and improves the reliability of the results. The strength of *MITHrIL* lies in the enrichment of pathways with information regarding microRNAs, post-transcriptional regulatory elements whose addition is clearly essential to the greater meaningfulness of the results. Our method, starting from expression values of genes and/or microRNAs, returns a list of pathways sorted according to the degree of their de-regulation, together with the corresponding statistical significance (p-values), and a predicted degree of alteration for each endpoint (a pathway node whose alteration, based on current knowledge, affects the phenotype in a specific way).

To evaluate our algorithm, we used the decoy pathway methodology introduced in Vaske et al. [28] on expression datasets obtained from *The Cancer Genome Atlas*. We showed that adding information on the otherwise missing regulatory elements proves to be pivotal in improving the reliability of pathway analysis methodologies. As further evidence of the reliability of our pathway impact analysis method, we employed our algorithm for the classification of phenotypes. The results highlight the ability of our methodology to strongly reduce the dimensionality of the data while maintaining a very high classification quality.

## Results

**Biological soundness.** We compared our methodology with *PARADIGM* [28], *SPIA* [27] and Micrographite [30] by employing the technique defined in Vaske et al. 2010. The aim is to establish whether the ranking computed with a pathway analysis algorithm is biologically significant. This is achieved by defining random pathways (called decoy pathways) with the same topology as the real ones but randomly selected nodes. All pathways are then evaluated by each algorithm, estimating the ability of each method to properly separate decoy pathways from real ones by means of a receiver operating characteristic (ROC) curve. In principle, a method that can correctly distinguish real pathways from decoys should yield biologically significant results.

We performed comparisons between *MITHrIL, SPIA, PARADIGM and Micrographite* on a set of selected cancer types (see Table 1). Such a comparison allowed us, by ranking the datasets according to performance, to identify the single cancer type in which our algorithm had the lowest quality results, namely, Lung squamous cell carcinoma (LUSC).

The results of the four methodologies were ranked as follows: *PARADIGM* according to the average number of significant scores, as described in [28]; *SPIA* according to the adjusted p-value as obtained through their software implementation; Micrographite according to the pathway prioritization phase; *MITHrIL* according to the adjusted p-value and the accumulator. More precisely, in MITHril, all the results are sorted first by p-value and, thereafter, in the presence of equal p-values their corresponding accumulator is taken into account to determine their order. In Supplementary Figures 1-3, we present the results of the detailed comparison in each TCGA dataset. Out analysis clearly shows that *MITHrIL* gives the best performances. As further proof of the goodness of our methodology, we computed the average area under each ROC curve (AUC). The results were summarized in Figure 1 (more details can be found in Supplementary Table 1). The four boxes in the figure represent the AUC variability range for the four compared methodologies, respectively.

**Prediction of cancer types.** We also evaluated our algorithm by assessing its performances in terms of capability to predict the cancer type. To do this, we elected to train the *PAMr* [31] classification algorithm and evaluated its performance by means of a 10-fold cross validation (CV) procedure. *PAMr* is an approach devised to predict cancer class from gene expression profiling, based on an enhancement of the *nearest shrunken centroid* classifier. The algorithm is able to identify subsets of genes that best characterize each class. The technique is general and can be used in many other classification problems. The CV procedure takes as input all the feature profiles of each patient, and divides them into 10 subsets, by balancing the elements of each class in each subset. A subset is, then, removed (test set), and the classifier is trained on the remaining nine sets (training set), in order to prioritize and select the features. Each profile in the test set is then classified, and the results are used to estimate the error. The methodology is repeated so that each subset is used once as the test set. The CV procedure was designed in order to remove overfitting and overestimation of the results.

A reference classification was thus established by applying such a procedure to the Log-Fold-Change of differentially expressed genes of our cancer cases. The rationale behind such a choice is to show that pathway perturbation, which takes into account network structure, increases the biologically soundness of the results with respect to a plain Log-Fold-Change-based approach, widely used as a gold standard. First we computed all differentially expressed genes for each tumor type, obtaining a total of 17,326 genes that appear to be de-regulated in at least one disease. Next, we calculated their Log-Fold-Change in each sample, trained a classifier and verified its performance using the above described CV procedure. The results (Table 2) demonstrate that such a classification is quite reliable since it yields a very small error. Notice that Micrographite is not able to compute pathway ranking for a single sample, therefore we did not perform any classification using such a method. Hence, we ran MITHrIL, *SPIA* and *PARADIGM* on all samples of our set of selected cancer types, and trained three classification models using total accumulation scores. As before, we performed a 10-fold cross validation and evaluated errors in each class (Table 2). Furthermore, leveraging the ability of MITHrIL and PARADIGM to return the perturbation for each of 3,165 pathway endpoints, we trained additional classifiers based on such values. Since SPIA computes pathway-level statistic by means of a linear equation system, it could not return perturbations of endpoints. Therefore we elected to use MITHrIL without miRNAs to establish the classification performances of endpoints, when such elements are missing.

Our analysis clearly shows that performances are considerably improved over reference classification, by taking into account endpoint perturbations (Table 2). Moreover, we can notice a significant dimensionality reduction of our data, since by using perturbation of pathway endpoints, computed by means of Equation 1 (see the section Materials and Method), we are able to train *PAMR* on about 3165 genes (18% of the number of differentially expressed ones).

Table 2 reports also the classifications based on MITHrIL pathway accumulators. We recall that accumulator summarizes, with a single value, the general perturbation we observe within a pathway. Hence, as a further effect this yields a stronger dimensionality reduction. Although we notice a slight increase in misclassification error, compared to reference classification, it is important to highlight that we were able to reduce to 237 the number of features on which *PAMR* classifier was trained. Pathway Accumulators were computed according to Equation 5. The last two columns of Table 2 report the classification performances obtained by *SPIA* accumulators and PARADIGM scores. All of this shows that the addition of miRNA information is crucial in order to obtain more reliable results. Notice that we cannot deduce any information about the performances of PARADIGM extended with microRNAs, since no implementation with such a knowledge is available, and this goes beyond the scope of our paper.

To further highlight the biological relevance of endpoints, we performed a set of experiments with randomly selected nodes within pathways (see third and fourth columns of Table 2). The results show that the choice of endpoints is reasonable, since endpoints synthetize perturbations of upstream nodes.

**Coherence of the prediction of pathway nodes state.** As a further validation our methodology, we chose to verify the percentage of endpoints for which a coherent prediction of the deregulation is obtained. Initially, we applied MITHrIL with and without miRNAs to estimate the perturbations for each endpoint of each sample (excluding the expression values of the endpoints in order to avoid introducing a bias in our results). Subsequently, we computed the percentage of endpoints for which the sign of perturbation agrees with that of the log-Fold Change. This validation estimates the reliability of the predictions of our method and the importance of the addition of miRNA knowledge to our model. The results (Figure 2) highlight that plugging quantitative information on miRNAs is crucial to establish a far more comprehensive and meaningful estimation of phenotype activity. Therefore, using perturbation without miRNAs could be misleading.

## Discussion

In the last decade, miRNAs have ever more revealed to be crucial in the modulation of numerous cellular pathways via the exertion of their important regulatory function when targeting key genes. Since the first connection between miRNAs and cancer was made in 2002 [32], miRNA deregulation has been proven to be indeed at the root of the pathogenesis of all cancers [33]. It suffices to consider, for instance, how the deregulation of even a single miRNA is capable to cause cancer, as in the case of miR-155 which has been shown to be responsible for the onset of Acute Lyphoblastic Leukemia/high-grade lymphoma in mouse [34]. Additionally, the predominant roles played by miRs 21, 221 and 222 in several cancer types prove the importance these small RNA molecules have in tumor pathogenesis and progression, while also being a determining factor in drug resistance [33]. In light of this and many other evidences discovered in recent years, the integration of miRNA expression when evaluating cancer pathway perturbation has become of utmost importance. The proper consideration of the crucial effects yielded by the action of these small non-coding RNA molecules on overall gene expression indeed contributes to a more comprehensive depiction of the biological reality, providing a more accurate means for pathway assessment and phenotype categorization. In fact, given the very important biological role played by miRNAs, integrating their evaluation can greatly help in the discernment of even fine changes in the cellular gene expression profile, which could make the difference between a normal and abnormal phenotypes, already at disease onset.

Here we presented a novel knowledge base-driven pathway analysis methodology called *MITHrIL*. By enriching *KEGG* pathways with experimentally validated interactions between genes and miRNAs, *MITHrIL* is capable to clearly improve the reliability of pathway-based analysis of phenotypes.

Through the enrichment with miRNA information, *MITHrIL* can greatly improve predictions over *SPIA, PARADIGM, and Micrographite*. In fact, while the other methodologies cannot properly distinguish between decoy pathways and real ones, MITHrIL is capable of obtaining much better results. Even our worst case had superior results than our three competitors. From a biological standpoint, the ability to distinguish decoy pathways from real ones addresses the fundamental necessity to be able to properly interpret the actual cellular mechanisms as possessing a biological criterion which is crucial to the life of the cell and not the result of random phenomena.

Therefore, we focused our analysis on the ability of our methodology to synthesize information gained from gene expression data and thus provide novel biological clues. For this purpose, by using the *PAMR* algorithm, we performed different types of classification, taking as reference the results arising from the classification based directly on Log-Fold-Changes. Our findings demonstrated the capability of *MITHrIL* to synthesize biological information contained in the data, while yielding high classification accuracy. Furthermore, *MITHrIL* greatly reduces the dimensionality of data of about 73 times compared to a naive Log-Fold-Change based method. This significant dimensionality reduction may also make the analysis more accurate since it could reduce the noise that can be introduced by the technologies used to gather expression data from samples. Furthermore, when knowing the phenotype that is being analyzed, it is possible to further reduce the number of dimensions by focusing only on those pathways that are known to be somehow involved in it.

Compared to *SPIA*, MITHrIL can also return the perturbation computed for pathway endpoints, whose subsequent analysis can lead to important additional insights about the biology underlying the phenomena under study. Indeed, the proper evaluation of pathway endpoints can contribute to a far more accurate phenotype assessment, as a more detailed diversification between pathological phenotypes at the pathway level is reflected more at the endpoints rather than in any other node of the pathway network. This allows to more effectively distinguish pathologies sharing even a very similar set of deregulated genes (as you could more easily distinguish similar yet different trees more easily by confronting their leaves rather than their roots). By leveraging on endpoint perturbation, we are also able to greatly reduce the misclassification error, although we are able to reduce data dimensionality by only 5 times. This allows us to stress the fact that gene perturbations are capable to discriminate among the pathological classes of our data.

Leveraging on the potential provided by miRNA enrichment in pathway analysis, MITHrIL represents a bioinformatic resource capable of a far more accurate evaluation of pathway deregulation in cancer. This can provide a decisive contribution to cancer research in terms of directing researchers more effectively, reducing costs and time requirements. Specifically, MITHrIL can contribute to an earlier diagnosis, an early and more accurate drug resistance assessment, as well as to more precise prognosis in terms of predicting future disease development.

Nevertheless, our results can be further improved given that pathways are still incomplete, thus potentially resulting in partial or erroneous conclusions. Future development in pathway analysis methodologies should take into account additional regulatory elements, such as long non-coding RNA (lncRNAs), along with epigenetic information, such as methylation patterns, variants, or copy number variants. Mutations could be exploited by considering their impact on the modulation of $\beta$ function (see the section Materials and Method for the definition of $\beta$), for example by assessing the difference in interaction strength by means of free energy. We could also define the $\beta$ function by evaluating the correlation between patient expression profiles and corresponding phenotype.

## Materials and Methods

### Pathway Enrichment Outline

Our methodology distinguishes itself from other pathway analysis techniques primarily for the use of *KEGG* [4, 5] pathways enriched with microRNAs (miRNAs) and their interactions with genes.

In order to achieve this, we downloaded all validated inhibition interactions between miRNA and targets from *miRTarBase* [35] and *miRecords* [36]. We also obtained interactions between transcription factors (TFs) and miRNAs from *TransmiR* [37]. By taking into account TFs activating miRNA genes we are able to increase the knowledge stored within each pathway. We then standardized all identifiers in their respective databases to avoid duplicates. The mapping of miRNA identifiers was performed by using *miRBase release 20* [38-42] as reference database. For each target, we performed a twofold mapping procedure: firstly, each gene identifier has been converted to its Entrez one; then, by taking advantage of *KEGG REST API*, we mapped each *Entrez Id* to the corresponding *KEGG Id*. This standardized list of interactions was, lastly, filtered to remove all duplicates. Such a procedure allowed us to build a knowledge base of 10,537 experimentally validated interactions between 385 miRNAs and 3,080 genes.

Pathway enrichment was performed by defining a new type of nodes representing miRNAs in the pathway notation, along with two types of directed edges, for miRNA-target inhibition interactions and TF-miRNA interactions, respectively. The enrichment is thus performed automatically by adding to each pathway only miRNAs that interact with at least one element within it.

Finally, in order to acquire information on which endpoints are contained in each pathway, we employed a depth-first search algorithm (DFS) [43] to automatically mark which genes are located at the end of the chains of reactions in each pathway. The search for endpoints in a pathway starts from a random node. The DFS algorithm follows the interactions down to the nodes from which no other one can be reached (putative endpoints). The procedure is repeated until all nodes have been analyzed. Putative endpoints are, then, manually screened to determine if they are associated with phenotypic changes as stated on the KEGG database. Only the latter are taken as pathway endpoint. An example of endpoints is reported in Supplementary Figure 4.

### Algorithm

Our methodology consists in an extension of Draghici et al. [26] and Tarca et al. [27]. It requires a case/control expression data set from which statistically differentially expressed features have been extracted (genes, miRNAs, or both). For such elements, the computation of their Log-Fold-Change is also needed. Starting from such information, MITHrIL computes, for each gene in a pathway, a *Perturbation Factor* (*PF*), which is an estimate of how much its activity is altered considering its expression and 1-neighborhood. Positive (negative) values of *PF* indicate that the gene is likely activated (inhibited). By appropriately combining each *PF* of a pathway, our algorithm is, therefore, able to calculate an *Impact Factor* (*IF*) and an *Accumulator* (*Acc*). The *IF* of a pathway is a metric expressing how important are the changes detected in the pathway, the greater the value, the most significant are the changes. The *Acc* indicates the total level of perturbation in the pathway and the general tendency of its genes: positive *Acc* values indicate a majority of activated genes (or inhibited miRNAs), while negative ones corresponds to an abundance of inhibited genes (or activated miRNAs). To the *Acc* is also assigned a p-value which is an estimate of the probability of getting such accumulator by chance. Finally, by applying the [44] method, we estimate the false discovery rate and p-values are adjusted on multiple hypotheses.

More precisely, let $n$ be a node in pathway $P_i$. Its *perturbation factor*, $PF(n, P_i)$ can be defined as:

$$PF(n, P_i) = \Delta E(n) + \sum_{u \in U(n, P_i)} \frac{\beta(u, n)}{\sum_{d \in D(u, P_i)} |\beta(u, d)|} \cdot PF(u, P_i), \tag{1}$$

where $\Delta E(n)$ is the Log-Fold-Change computed for the node $n$, $U(n, P_i)$ and $D(n, P_i)$ are the set of upstream and downstream nodes of $n$ in pathway $P_i$ respectively, and $\beta(u, n)$ is a function that indicates the strength and type of interaction between genes $u$ and $n$. In particular, negative values of $\beta$ indicate an inhibitory effect, while positive values an activating one. To ensure that the perturbation coming from an upstream node is divided to its downstream ones, proportionally to the strength of their interactions, without altering the total perturbation, a normalization is applied dividing by the sum of the weights absolute values. By exploiting the methodology described in Draghici et al. [26] we compute an *impact factor*, $IF(P_i)$, which reflects the importance of the changes observed in a pathway, as:

$$IF(P_i) = \log\left(\frac{1}{p(P_i)}\right) + \frac{\sum_{n \in P_i} |PF(n, P_i)|}{\overline{|\Delta E|} \cdot N_{de}(P_i)}, \qquad (2)$$

where $p(P_i)$ is the probability, calculated using an hyper-geometric distribution, of obtaining a number of differentially expressed nodes at least equal to the observed one in $P_i$; $\overline{|\Delta E|}$ is the mean Log-Fold-Change in $P_i$; finally, $N_{de}(P_i)$ represents the number of differentially expressed nodes in the pathway.

Our methodology takes also advantage of the *accumulation* (or *accumulator*) as described by Tarca et al. [27]. Such a methodology has been revised to take into account the addition of miRNAs. In order to do so, first we need to compute two partial accumulators, $Acc_{mir}(P_i)$ and $Acc_{gene}(P_i)$, which take into account the perturbation, respectively, of miRNAs and genes:

$$Acc_{mir}(P_i) = \sum_{m \in P_i^m} [PF(m, P_i) - \Delta E(m)], \qquad (3)$$

$$Acc_{gene}(P_i) = \sum_{g \in P_i^g} [PF(g, P_i) - \Delta E(g)], \qquad (4)$$

where $P_i^m$ and $P_i^g$ are the sets of miRNAs and genes present in $P_i$, respectively.

Therefore, in equations 3 and 4, we sum the perturbations of all miRNAs ($P_i^m$) and genes ($P_i^g$) in pathway $P_i$, addressing the dominant effect of the expression change in the PF computation by subtracting such values. We can now compute total perturbation *accumulation*, $Acc(P_i)$, which measures whether the pathway is likely activated or inhibited. The introduction of miRNAs in our model addresses the necessity to take into account the fact that an increased (decreased) expression of such elements results in an inhibition (activation) of the pathway. We compute $Acc(P_i)$ as:

$$Acc(P_i) = Acc_{gene}(P_i) - Acc_{mir}(P_i) - E[Acc(P_i)], \qquad (5)$$

where $E[Acc(P_i)]$ is an estimate of the expected value of the distribution of all accumulators computed for pathway $P_i$, as explained below.

P-value estimation is then performed by combining the Z-scores, computed through an inverse Standardized Normal distribution, associated to two probabilistic terms: the first is the probability of obtaining by chance a number of differentially expressed genes in the pathway at least equal to the observed one, while the second consists in the probability of observing by chance an accumulator higher than the computed one. The first term corresponds to $p(P_i)$ introduced in equation 2. The second term, instead, has to be estimated through a permutation test. In such a test, we assign, to a random group of genes in the pathway in question, a Log-Fold-Change selected randomly from the input ones, so as to compute a random accumulator. The procedure is repeated several times and the final probability is estimated as the ratio between the number of random accumulators greater than $Acc(P_i)$ and the number of repetitions performed. In our experiments, the repetitions were set to 2000 in order to obtain maximum precision up to two decimal places.

At this stage we are also able to estimate expected value $E[Acc(P_i)]$ as the median value of the random accumulators.

Therefore, the final result of our algorithm consists of a list of pathways along with their impact factor, accumulator and adjusted p-values. Such list is sorted by p-value and *Acc*.

Expression Data Sources

To perform a comprehensive test of our algorithm, we exploited expression data provided by *The Cancer Genome Atlas* (beginning of 2014). We downloaded all patient expression profiles of genes (RNASeqV2 obtained through platforms Illumina Genome Analyzer and Illumina HiSeq) and miRNAs (miRNASeq obtained through platforms Illumina Genome Analyzer and Illumina HiSeq). The initial dataset was then filtered by removing all patients for which one of the two types of expression was unavailable. We then eliminated all tumor samples for which no healthy controls were available. By applying such a procedure, we built a dataset of 3,053 expression profiles (2,721 case samples and 332 control samples) of patients affected by 10 distinct tumor pathologies (see Table 1 for more details). Case samples were further divided by disease stage.

To run our algorithm, we performed a differentially expressed genes analysis by using the *RNASeq* pipeline based on *Limma* [45]. The expression matrices for each disease were firstly normalized by using the *Voom* algorithm [46], then a linear model was trained with *Limma* and differentially expressed genes for each stage of the disease were extracted along with their Log-Fold-Change. In our analysis we considered as differentially expressed only those genes for which an adjusted p-value was lower than 0.01 as computed by *Limma*.

In order to correctly ascertain *PARADIGM* performance, for each tumor sample we also downloaded and processed copy number variation (CNV) as shown in Vaske et al. 2010 [28].

Performance Assessment

To compare our algorithm with other methodologies, *PARADIGM* [28], *SPIA* [27] and Micrographite [30], we used the decoy pathway technique introduced in Vaske et al. 2010 [28]. For each pathway in our internal database, we built a decoy one obtained by maintaining the same structure and substituting each gene (or miRNA) with one randomly chosen from the set of all possible genes. As in Vaske et al. 2010 [28], all the complexes and abstract processes were kept unchanged. After the execution of the three algorithms, the pathways were classified by each method and the fraction of real pathways versus the total number of pathways considered was computed. The higher the fraction of real pathways, the better the ability of an algorithm to extract biologically sound results. Lastly, to achieve a fair comparison with *SPIA*, we chose the same $\beta$ function as Tarca et al. 2009 [27]: $\beta(u,n) = 1$ for all interactions that increase node expression level, $\beta(u,n) = -1$ for those that have the effect of decreasing node expression level, $\beta(u,g) = 0$ for irrelevant ones. However, the $\beta$ function introduces a huge concealed potential in MITHrIL, which paves the way for possible future extensions.

# Acknowledgements

The results shown here are in whole or part based upon data generated by the TCGA Research Network: http://cancergenome.nih.gov/. V. D. was supported by Italian Foundation for Cancer Research (FIRC) (16572).

# Conflicts of Interest

The authors declare no conflict of interest.

## Tables

| Code | Cancer Type | Control Samples | Case Samples | Case Samples Categories |
|---|---|---|---|---|
| BLCA | Bladder Urothelial Carcinoma | 19 | 193 | Stage I, II, III, IV |
| BRCA | Breast invasive carcinoma | 86 | 642 | Stage I, II, III, IV, X |
| COAD | Colon adenocarcinoma | 8 | 389 | Stage I, II, III, IV |
| KICH | Kidney Chromophobe | 25 | 66 | Stage I, II, III, IV |
| KIRC | Kidney renal clear cell carcinoma | 71 | 224 | Stage I, II, III, IV |
| LUAD | Lung adenocarcinoma | 19 | 388 | Stage I, II, III, IV |
| LUSC | Lung squamous cell carcinoma | 37 | 247 | Stage I, II, III, IV |
| PRAD | Prostate adenocarcinoma | 50 | 191 | Category 6, 7, 8, 9, 10 |
| READ | Rectum adenocarcinoma | 3 | 150 | Stage I, II, III, IV |
| UCEC | Uterine Corpus Endometrial Carcinoma | 14 | 231 | Stage I, II, III, IV |
|  | **All Samples** | **332** | **2721** |  |

Table 1 List of cancer types extracted from The Cancer Genome Atlas (TCGA) with their codes, number of case and control samples, and Subcategories.

| Data | Log-FC | Perturb. Of Random Nodes | | Endpoints | | | Pathway-level Statistics | | |
|---|---|---|---|---|---|---|---|---|---|
|  |  | MITHrIL | MITHrIL no miRNA | MITHrIL | MITHrIL no miRNA | PARADIGM | MITHrIL Acc. | SPIA Acc. | PARADIGM Scores |
| **BLCA** | 3.11% | 9.59% | 6.58% | **1.55%** | 2.60% | 2.08% | 12.95% | 49.74% | 82.38% |
| **BRCA** | 1.86% | 2.12% | 3.97% | **1.09%** | 2.00% | 2.34% | 13.08% | 8.25% | 73.05% |
| **COAD** | 2.31% | **0.00%** | 7.81% | 0.00% | 0.00% | 3.10% | 0.77% | **0.00%** | 32.90% |
| **KICH** | 3.03% | 1.67% | 3.03% | **0.00%** | 0.00% | 3.03% | 4.54% | 3.03% | 31.81% |
| **KIRC** | 3.12% | 2.68% | 2.77% | **1.79%** | 2.68% | 3.13% | 5.80% | 2.67% | 35.26% |
| **LUAD** | 4.89% | **0.03%** | 8.61% | 1.80% | 2.06% | 5.41% | 4.38% | 2.83% | 64.43% |
| **LUSC** | 6.07% | 1.78% | 6.92% | **1.21%** | 2.02% | 6.91% | 5.26% | 4.04% | 71.54% |
| **PRAD** | **0.00%** | 0.37% | 1.26% | 0.00% | 0.00% | 0.52% | 2.61% | 30.89% | 18.94% |
| **READ** | 3.33% | **0.00%** | 9.40% | 0.00% | 0.00% | 4.00% | 0.66% | **0.00%** | 96.66% |
| **UCEC** | 1.73% | 1.39% | 1.13% | **0.00%** | 0.43% | 0.09% | 4.32% | 1.29% | 46.32% |
| **Total** | 2.90% | 1.75% | 5.38% | **0.90%** | 1.50% | 3.20% | 6.40% | 8.80% | 57.60% |

Table 2 Classification results of tumor samples in our dataset obtained training PAMR algorithm by means of Log-Fold-Change, SPIA total accumulation, Paradigm scores, MITHrIL accumulators, and MITHrIL endpoint perturbations. Each element in the table corresponds to the classification error for a specific cancer type using one algorithm. Despite the reference classification based on Log-Fold-Change yields a low average error (2.90%), the employment of perturbations computed for each endpoint provides a significant improvement in the classification accuracy.

# Figures

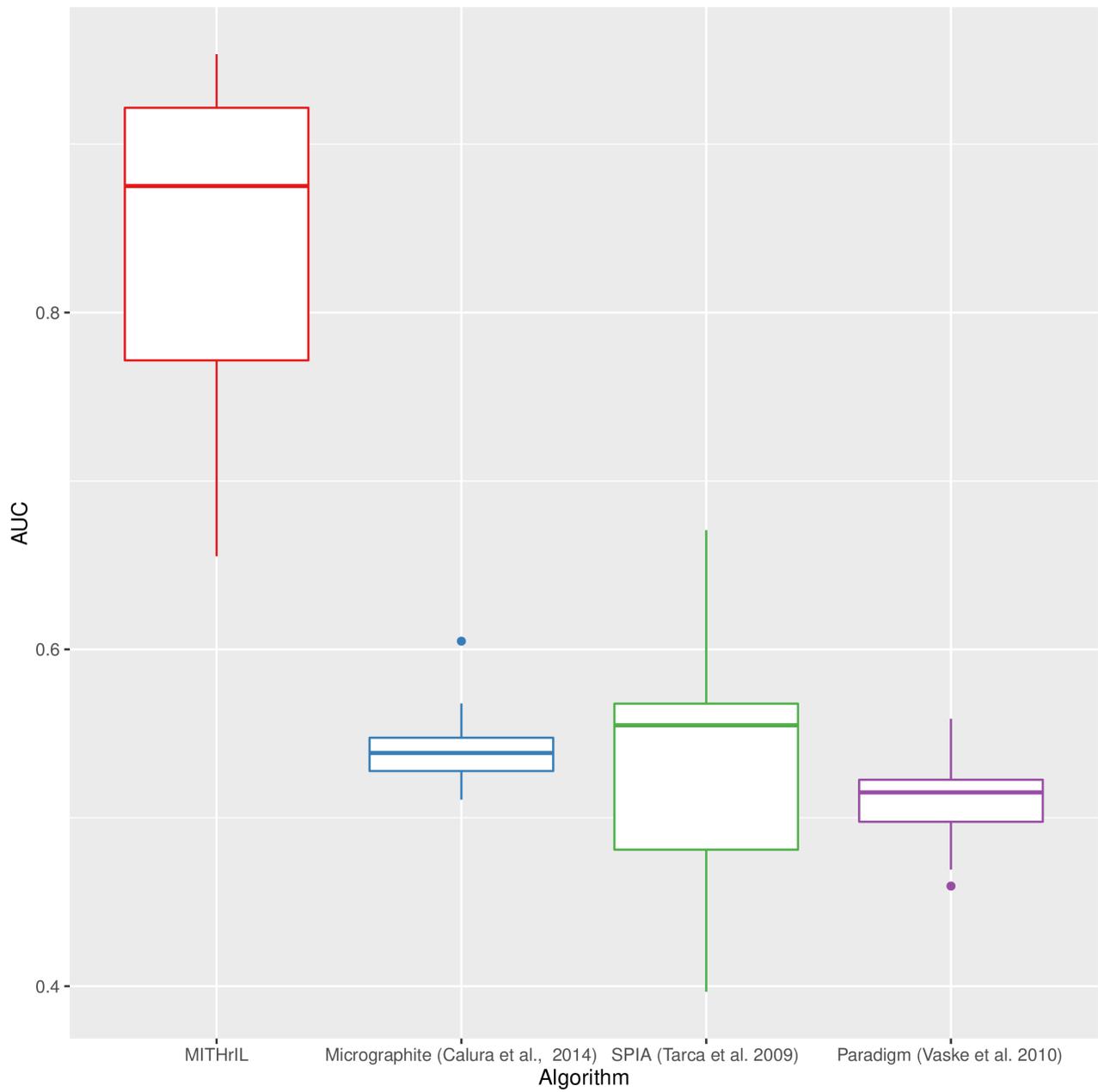

*Figure 1 Performances comparison between MITHrIL, SPIA, and PARADIGM by means of the average area under the ROC curves. Each box in the figure represents the variability range of AUC values for a specific methodology.*

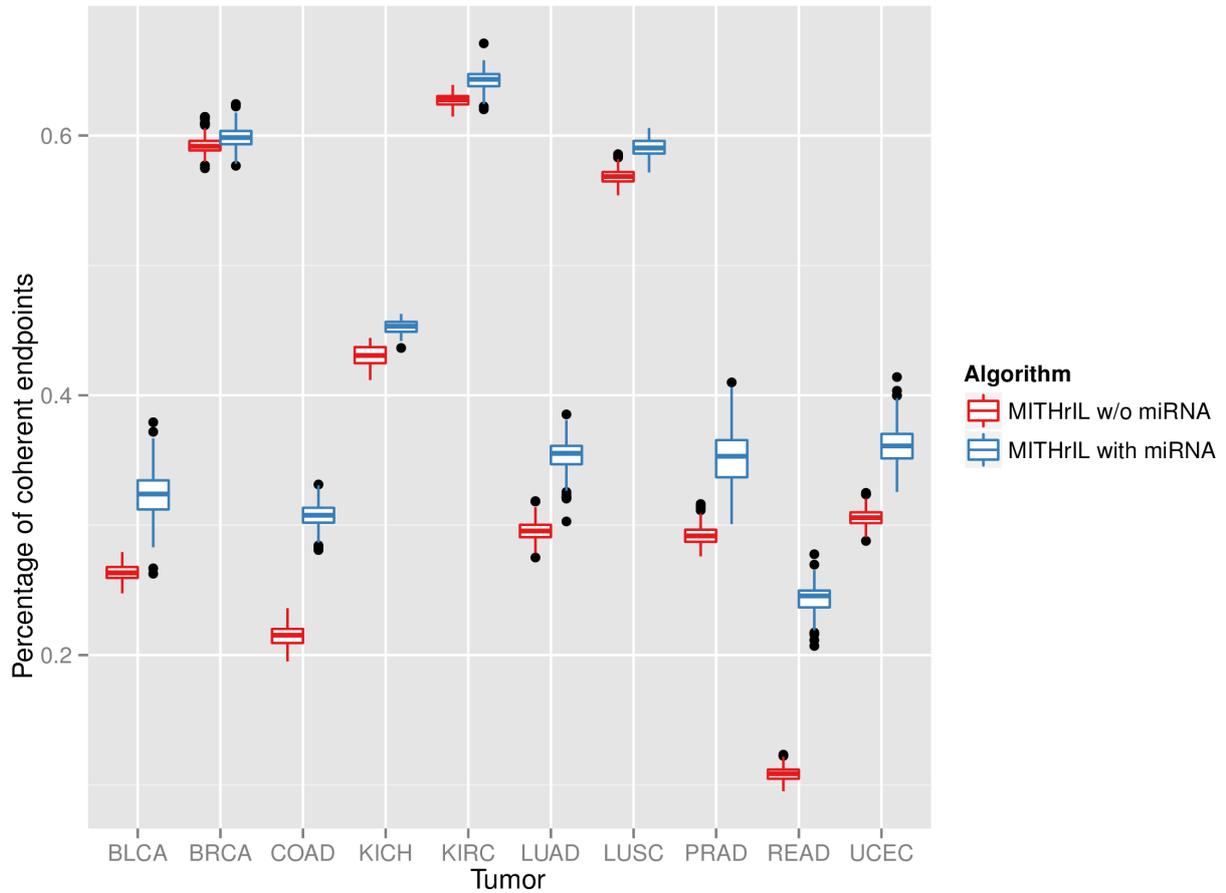

Figure 2 Significance of the addition of miRNA in our model by means of a comparison of the percentages of correctly predicted endpoints for each sample between our method with and without miRNAs. Each box in the figure represents the variability range of the percentage of correctly predicted endpoints for the patients of a specific tumor type. A prediction is correct when the deregulation observed in the original data correspond to the one inferred by our algorithm. Namely, the sign of an endpoint log-Fold-Change corresponds to the sign of its perturbation value.